\documentclass[12pt]{article}

\usepackage{scicite}

\usepackage{times}
\usepackage{color}

\usepackage{graphicx}

\newenvironment{sciabstract}{%
\begin{quote} \bf}
{\end{quote}}

\title{An extremely metal-deficient globular cluster in the Andromeda Galaxy}

\author
{S{\o}ren S.\ Larsen,$^{1\ast}$ Aaron J.\ Romanowsky,$^{2,4}$ Jean P.\ Brodie,$^{3,4}$ Asher Wasserman$^{4}$\\
\\
\normalsize{$^{1}$Department of Astrophysics/Institute of Mathematics, Astrophysics and Particle Physics,} \\
\normalsize{Radboud University, 6500 GL Nijmegen, The Netherlands}\\
\normalsize{$^{2}$Department of Physics and Astronomy, San Jos{\'e} State University,}\\
\normalsize{San Jose, CA 95192, USA}\\
\normalsize{$^{3}$Centre for Astrophysics and Supercomputing,}\\
\normalsize{Swinburne University of Technology, Hawthorn, VIC 3122, Australia}\\
\normalsize{$^{4}$University of California Observatories, }\\
\normalsize{ University of California, Santa Cruz, CA 95064, USA}\\
\normalsize{$^\ast$To whom correspondence should be addressed; E-mail:  s.larsen@astro.ru.nl.}
}

\date{}

\begin{document} 


\baselineskip24pt


\maketitle


\begin{sciabstract}
Globular clusters (GCs) are dense, gravitationally bound systems of thousands
to millions of stars. They are preferentially associated with the oldest
components of galaxies, and measurements of their composition can therefore
provide insight into the build-up of the chemical elements in galaxies in the
early Universe.
We report a massive GC in the Andromeda Galaxy (M31) that
is extremely depleted in heavy elements. Its iron abundance is about 800 times
lower than that of the Sun, and about three times lower than in the most iron-poor
GCs previously known.  It is also strongly depleted in magnesium. These
measurements challenge the notion of a metallicity floor for GCs and
theoretical expectations that massive GCs could not have formed at such
low metallicities.
\end{sciabstract}

Globular clusters (GCs) are roughly spherical agglomerations of thousands to millions of stars, bound by their mutual gravity, and have central densities that can exceed $10^6$ solar masses per cubic parsec ($M_\odot \, \mathrm{pc}^{-3}$) \cite{Baumgardt2018}. 
GCs formed early in the history of the Universe and therefore record the early stages of galaxy formation and evolution. 
The nearest neighboring spiral galaxy, the Andromeda Galaxy, also known as Messier 31 (M31), has a system of GCs that align spatially and kinematically with stars in the outer parts of the galaxy. The GCs in the outer parts of M31 appear to belong to at least two kinematically distinct subsystems that were accreted separately  \cite{Mackey2019}. 

The GC systems in most galaxies are dominated by clusters with low abundances of elements heavier than hydrogen and helium (``metals'') relative to the composition of the Sun. However, there appears to be a deficit of GCs at the very lowest metal abundances (``metallicities'') \cite{Carney1996}. 
The most metal-poor GCs in the Milky Way have iron abundances of $\mathrm{[Fe/H]}\approx-2.5$ \cite{Simpson2018}, where square brackets denote the abundance ratios of the elements, relative to the solar photospheric composition, on a logarithmic scale. The number of
iron atoms per hydrogen atom in the most metal-poor GCs is thus about 300 times lower than in the Sun.
The notion of a metallicity floor for GCs at $\mathrm{[Fe/H]}=-2.5$ is  
supported by observations of GCs in several external galaxies \cite{Beasley2019}, and various explanations have been suggested. 
The correlation between mass and metallicity for galaxies in the early Universe might set a minimum metallicity for formation of GCs that are sufficiently massive to survive until the present day, or the formation of massive GCs could be suppressed at low metallicities due to inefficient gas cooling \cite{Carney1996,Choksi2018,Beasley2019,Kruijssen2019}.
A metallicity floor for GCs would thus have implications for cluster- and star formation and for the build-up of metals in galaxies in the early Universe.

Because the metallicity distributions of both GCs and individual stars decline steeply towards low metallicities and are poorly constrained, it remains unclear how statistically significant the proposed metallicity floor is. In M31, three clusters with metallicities that may fall in the range $-2.8 < \mathrm{[Fe/H]} < -2.5$ are known \cite{Caldwell2011}, but the uncertainties are large (0.3-0.4 dex) and the metallicities may lie well above the floor. Similarly, three GCs in the Sombrero galaxy may have metallicities below $\mathrm{[Fe/H]}=-2.5$ \cite{AlvesBrito2011}, but the uncertainties on the spectroscopic measurements are large and the red colors of these clusters suggest higher metallicities. 
 
We investigate the globular cluster RBC EXT8 (hereafter EXT8) in M31, located at right ascension 00$^h$53$^m$14$^s$.53, declination $+$41$^\circ$33$^\prime$24$^{\prime\prime}$.5 (J2000 equinox)
according to the Revised Bologna Catalogue 
\cite{Galleti2004}.  
From a kinematic analysis \cite{Mackey2019}, EXT8 belongs to 
the smoothly distributed component of the M31 halo, and lies at a projected distance of 27 kpc from the galaxy center. 
Figure~1 shows a color-magnitude diagram for GCs in M31 \cite{Peacock2010}. With an apparent magnitude in the $g$-band of $g=15.87$, EXT8 is among the brighter GCs, and its integrated light color with respect to the $u$-band ($u-g=1.11$) is less red than most of the other GCs, suggesting a low metallicity. 
Previous low-resolution spectroscopy yielded an age $\geq 8$~Gyr and [Fe/H] between $-2.8$ and $-2.0$  
\cite{Fan2011,Chen2016}.

We obtained a spectrum of the integrated light of EXT8 with the High-Resolution Echelle Spectrometer (HIRES)  \cite{Vogt1994} on the Keck~I telescope on 25 Oct 2019. Given EXT8's high brightness and compact size, a total integration time of 2400 s was sufficient to obtain a signal-to-noise ratio of about 200 per \AA\ near the Mg~{\sc i} triplet at 5170 \AA .
We used a slit width of 1.15$^{\prime\prime}$ which gave a nominal spectral resolving power $R\equiv\lambda/\Delta\lambda\approx37000$ for wavelength $\lambda$ and width $\Delta \lambda$ of a spectral resolution element. The observations covered a spectral range of 3840-8060~\AA .

Figure~2 shows the H$\beta$ lines in the spectra of EXT8 and Messier~15 (M15) for comparison, the latter being one of the most metal-poor GCs in the Milky Way \cite{Harris1996}. 
The spectrum of M15 was obtained with the Ultraviolet and Visual Echelle Spectrograph (UVES) on the Very Large Telescope and has a spectral resolving power similar to that of the EXT8 spectrum \cite{Larsen2017}.
M15's velocity dispersion ($\sigma = 12.9\pm0.3$ km~s$^{-1}$) is similar to that of EXT8 ($\sigma=13.3\pm0.8$ km~s$^{-1}$), allowing direct comparison.
Absorption in the H$\beta$ line becomes stronger at younger ages and can be used as an age indicator in the spectra of GCs \cite{Worthey1994a}. 
While the blue color of EXT8 could, in principle, be caused by a younger age, 
Figure~2 shows no discernible difference in the strengths or shapes of the H$\beta$ lines in the two spectra, indicating that EXT8 is similarly old, so must be a metal-poor GC. 

Figure~3 shows two metallicity-sensitive features.
Figure 3A shows the Fe~\textsc{i} feature near 4957~\AA\ (actually a blend of several Fe~\textsc{i} lines, of which the two strongest are marked), which is much weaker in the spectrum of EXT8 than in M15. Figure 3B shows two of the three lines of the Mg~\textsc{i} triplet (Fraunhofer's $b$ feature) at 5167~\AA\ and 5173~\AA . The third line, at 5184~\AA, falls in the gap between the two detectors of UVES, but is included in the HIRES spectrum.
The Mg~\textsc{i} lines, as well as other lines visible in this region of the spectra, are much weaker in the EXT8 spectrum. 

To quantify these results, we analyzed the EXT8 spectrum using a spectral fitting technique used in previous studies of extragalactic GCs \cite{Larsen2017,Larsen2018}. 
Figure~3 shows the best-fitting model spectrum for M15 \cite{Larsen2017}, with an iron abundance of [Fe/H]=$-2.39\pm0.02$.  This model spectrum is based on a color-magnitude diagram (CMD) of M15 \cite{materials}.
We do not have spatially resolved data to empirically build a CMD for EXT8, and substituted it with stellar models \cite{Choi2016} with a metal fraction chosen to self-consistently match that derived from the spectral modeling. We found an iron abundance of [Fe/H]=$-2.91\pm0.04$ for EXT8 from model fitting of the wavelength range 4400-6200 \AA\ \cite{materials}. These model spectra are also shown in Figure~3. We tested the assumptions required for the input CMD and found that they do not substantially affect this measurement \cite{materials}. We conclude that EXT8 is about 0.5 dex more metal-poor than the value of [Fe/H]=$-2.39$ found for M15. This metallicity lies well below the metallicity floor suggested by previous studies.

At the signal-to-noise ratio of our EXT8 spectrum, most spectral features are weak because of the low metallicity. Nevertheless, our model fitting yielded abundances for several additional elements. 
From the Mg~\textsc{i}~$b$ lines we found a low magnesium abundance of 
[Mg/Fe]=$-0.35\pm0.05$. Magnesium is thus even more strongly depleted than iron, relative to the Sun.
The other Mg lines that are typically measured in integrated-light GC spectra are very weak (see Supplementary Text), but are generally consistent with the low magnesium abundance inferred from the $b$ lines: 
[Mg/Fe]=$-0.16^{+0.27}_{-0.58}$ (using Mg~\textsc{i} 4571 \AA), 
[Mg/Fe]=$-0.96^{+0.65}_{-\infty}$ (Mg~\textsc{i} 4703 \AA), and
[Mg/Fe]=$-0.35\pm0.25$ (Mg~\textsc{i} 5528 \AA).
The consistency between the magnesium abundances measured from the $b$ triplet and from the weaker lines is supported by prior analysis of M15, where the weaker lines yielded [Mg/Fe]$=+0.18\pm0.06$ \cite{Larsen2017}. From our model fitting of M15, we find an almost identical abundance of [Mg/Fe]$=+0.17\pm0.02$ (using the two Mg~$b$ lines included in the UVES spectrum).
Magnesium is among the elements thought to be produced via the $\alpha$ process \cite{Burbidge1957}.
For other $\alpha$-elements, we find average abundance ratios of [Si/Fe]=$+0.65\pm0.31$, [Ca/Fe]=$+0.35\pm0.07$, and  [Ti/Fe]=$+0.19\pm0.06$ for EXT8 \cite{materials}. Relative to iron, 
these elements are thus on average enhanced by roughly a factor of two compared to the composition of the solar photosphere. 

The enhanced abundances of silicon, calcium, and titanium are typical for metal-poor, old populations, which is usually attributed to enrichment of $\alpha$-elements dominated by core-collapse supernovae  \cite{Tinsley1979}. 
However, the very low magnesium abundance is not easily explained within this framework. It may be related to the phenomenon of multiple stellar populations in GCs, of which the outer halo cluster NGC~2419 is one of the most extreme cases in the Milky Way \cite{Cohen2012,Mucciarelli2012a}. 
In NGC~2419,  some individual stars have magnesium abundances as low as $\mathrm{[Mg/Fe]}=-1$, although more typical values for the Mg-depleted stars in NGC~2419 (which constitute about 40\% of the stars in this cluster) are $\mathrm{[Mg/Fe]}\approx-0.5$.
The distribution of $\mathrm{[Mg/Fe]}$ values within EXT8 is not constrained by our measurements, but the $\mathrm{[Mg/Fe]}$ value measured from the integrated light can be reproduced if the cluster contains two populations with $\mathrm{[Mg/Fe]}\approx-1.0$ and $\mathrm{[Mg/Fe]}\approx+0.3$ that each account for about half of the stars \cite{materials}. In this case, a larger difference between two populations is required in EXT8 than in NGC~2419. 

Further evidence for multiple populations in EXT8 comes from the abundance of sodium. From the Na~\textsc{i} resonance doublet at 5890, 5896 \AA\ (Fraunhofer's $D$ feature) we find sodium to be enhanced relative to scaled-solar composition, [Na/Fe]=$+0.23\pm0.07$, as is commonly observed in GCs \cite{Carretta2009a}.
However, the $D$ lines may be contaminated by absorption from interstellar gas along the line-of-sight towards EXT8. Other Na~\textsc{i} lines that are immune to this effect are very weak in the spectrum of EXT8, but from the doublet at 5683, 5688 \AA\ we measure [Na/Fe]=$+0.26^{+0.32}_{-\infty}$, which is consistent with the value inferred from the $D$ lines.

In NGC~2419, the Mg-poor stars are also enriched in K, reaching $\mathrm{[K/Fe]}\approx+1.5$ \cite{Cohen2012,Mucciarelli2012a}. For EXT8 we measure $\mathrm{[K/Fe]}=+0.67\pm0.15$, but this value may require correction downward by about 0.3 dex to account for our assumption of local thermodynamic equilibrium in the spectral modeling \cite{Takeda2002a}. This would then make the $\mathrm{[K/Fe]}$ ratio in EXT8 similar to that observed in metal-poor halo stars in the Milky Way and in M15 \cite{Takeda2009}. Hence, we find no evidence of a K-enriched population in  EXT8.

From the model fitting, we also determined the velocity broadening of the observed spectrum. 
The line-of-sight velocity dispersion, corrected for instrumental broadening, is $\sigma = 13.3\pm0.8$~km~s$^{-1}$. The half-light radius of EXT8 is 2.8 pc, leading to an estimated 
dynamical mass of $M_\mathrm{dyn} = (1.14\pm0.16)\times10^6 \, M_\odot$ \cite{materials}. For an absolute magnitude in the $V$-band of $M_V=-9.28$ \cite{Huxor2014}, 
the corresponding mass-to-light ratio in solar units is $M_\mathrm{dyn}/L_V = 2.6 \, M_\odot/L_{\odot, V}$, where $L_V$ and $L_{\odot,V}$ are the luminosities of EXT8 and the Sun in the $V$-band. EXT8 thus extends the trend for metal-poor GCs to have high $M/L$ values  \cite{Strader2011}.  A lower $M/L$ would be expected for a younger age, so the measured value is consistent with EXT8 being an old, metal-poor system. 

Figure~4 shows a comparison of EXT8 with previous integrated-light spectroscopy of Galactic and extragalactic GCs \cite{Larsen2017,Larsen2018}, along with literature data for individual stars in Galactic GCs and individual stars \cite{Pritzl2005,Venn2004}.  
EXT8 is an outlier in a [Mg/Fe] vs.\ [Fe/H] plot (Fig.~4A), being more metal-poor than other GCs and more magnesium poor than individual stars with similarly low iron abundances.  The GCs have a larger spread in [Mg/Fe] than the individual stars, with scatter towards lower magnesium abundances. This has previously been interpreted as a signature of multiple populations \cite{Colucci2009,Sakari2013}.  
When excluding magnesium, Figure 4B shows that EXT8 has an [$\alpha$/Fe] ratio (here defined as the average of [Ca/Fe] and [Ti/Fe]) that overlaps with those seen in individual metal-poor stars and in other GCs. 

Within the standard paradigm of hierarchical galaxy assembly, metal-poor GCs are expected to have formed in the early Universe in low-mass galaxies that merged to form larger galaxies \cite{Harris2006a,Choksi2018,Kruijssen2019}. The correlation between the mass and metallicity of galaxies therefore imprints a maximum mass for a GC that could form with a given metallicity. At $\mathrm{[Fe/H]}=-2.9$, the maximum mass is expected to be about $10^5 \, M_\odot$ \cite{Choksi2018,Kruijssen2019}. The existence of a possible remnant of a disrupted GC in the Milky Way with $\mathrm{[Fe/H]}=-2.7$ and an estimated mass below $10^5 \, M_\odot$  \cite{Wan2020} is consistent with this notion. 
However, clusters as massive and metal-poor as EXT8 should be extremely rare.
In a simulation of 10553 GCs with masses greater than $10^5 \, M_\odot$, only three ($\sim$0.03\%) had  $\mathrm{[Fe/H]} < -2.5$ and masses above $10^6 M_\odot$ \cite{Usher2018}, where we have converted the total metal fractions in \cite{Usher2018} to $\mathrm{[Fe/H]}$ values \cite{materials}. If half of the 400--500 GCs in M31 \cite{Galleti2004} have masses greater than $10^5\, M_\odot$, this would correspond to a probability of 6--7\% of finding a single GC as massive and metal-poor as EXT8.  

\bibliography{larsen}

\begin{thebibliography}{10}

\bibitem{Baumgardt2018}
H.~Baumgardt, M.~Hilker, {\it Mon. Not. R. Astron. Soc.\/} {\bf 478}, 1520
  (2018).

\bibitem{Mackey2019}
D.~Mackey, {\it et~al.\/}, {\it Nature\/} {\bf 574}, 69 (2019).

\bibitem{Carney1996}
B.~W. Carney, J.~B. Laird, D.~W. Latham, L.~A. Aguilar, {\it Astron. J.\/} {\bf
  112}, 668 (1996).

\bibitem{Simpson2018}
J.~D. Simpson, {\it Mon. Not. R. Astron. Soc.\/} {\bf 477}, 4565 (2018).

\bibitem{Beasley2019}
M.~A. Beasley, {\it et~al.\/}, {\it Mon. Not. R. Astron. Soc.\/} {\bf 487},
  1986 (2019).

\bibitem{Choksi2018}
N.~Choksi, O.~Y. Gnedin, H.~Li, {\it Mon. Not. R. Astron. Soc.\/} {\bf 480},
  2343 (2018).

\bibitem{Kruijssen2019}
J.~M.~D. Kruijssen, {\it Mon. Not. R. Astron. Soc.\/} {\bf 486}, L20 (2019).

\bibitem{Caldwell2011}
N.~Caldwell, R.~Schiavon, H.~Morrison, J.~A. Rose, P.~Harding, {\it Astron.
  J.\/} {\bf 141}, 61 (2011).

\bibitem{AlvesBrito2011}
A.~Alves-Brito, {\it et~al.\/}, {\it Mon. Not. R. Astron. Soc.\/} {\bf 417},
  1823 (2011).

\bibitem{Galleti2004}
S.~Galleti, L.~Federici, M.~Bellazzini, F.~F. Pecci, S.~Macrina, {\it Astron.
  Astrophys.\/} {\bf 416}, 917 (2004).

\bibitem{Peacock2010}
M.~B. Peacock, {\it et~al.\/}, {\it Mon. Not. R. Astron. Soc.\/} {\bf 402}, 803
  (2010).

\bibitem{Fan2011}
Z.~Fan, {\it et~al.\/}, {\it Research in Astronomy and Astrophysics\/} {\bf
  11}, 1298 (2011).

\bibitem{Chen2016}
B.~Chen, {\it et~al.\/}, {\it Astron. J.\/} {\bf 152}, 45 (2016).

\bibitem{Vogt1994}
S.~S. Vogt, {\it et~al.\/}, {\it Proc. SPIE\/}, D.~L. Crawford, E.~R. Craine,
  eds. (1994), vol. 2198, p. 362.

\bibitem{Harris1996}
W.~E. Harris, {\it Astron. J.\/} {\bf 112}, 1487 (1996).

\bibitem{Larsen2017}
S.~S. Larsen, J.~P. Brodie, J.~Strader, {\it Astron. Astrophys.\/} {\bf 601},
  A96 (2017).

\bibitem{Worthey1994a}
G.~Worthey, {\it Astrophys. J. Suppl. Ser.\/} {\bf 95}, 107 (1994).

\bibitem{Larsen2018}
S.~S. Larsen, J.~P. Brodie, A.~Wasserman, J.~Strader, {\it Astron.
  Astrophys.\/} {\bf 613}, A56 (2018).

\bibitem{materials}
{Materials and methods are available as supplementary materials}.

\bibitem{Choi2016}
J.~Choi, {\it et~al.\/}, {\it Astrophys. J.\/} {\bf 823}, 102 (2016).

\bibitem{Burbidge1957}
E.~Burbidge, G.~Burbidge, W.~Fowler, F.~Hoyle, {\it Rev. Mod. Phys.\/} {\bf
  29}, 547 (1957).

\bibitem{Tinsley1979}
B.~M. Tinsley, {\it Astrophys. J.\/} {\bf 229}, 1046 (1979).

\bibitem{Cohen2012}
J.~G. Cohen, E.~N. Kirby, {\it Astrophys. J.\/} {\bf 760}, 86 (2012).

\bibitem{Mucciarelli2012a}
A.~Mucciarelli, {\it et~al.\/}, {\it Mon. Not. R. Astron. Soc.\/} {\bf 426},
  2889 (2012).

\bibitem{Carretta2009a}
E.~Carretta, {\it et~al.\/}, {\it Astron. Astrophys.\/} {\bf 505}, 117 (2009).

\bibitem{Takeda2002a}
Y.~Takeda, G.~Zhao, Y.-Q. Chen, H.-M. Qiu, M.~Takada-Hidai, {\it Publ. Astron.
  Soc. Jpn.\/} {\bf 54}, 275 (2002).

\bibitem{Takeda2009}
Y.~Takeda, {\it et~al.\/}, {\it Publ. Astron. Soc. Jpn.\/} {\bf 61}, 563
  (2009).

\bibitem{Huxor2014}
A.~P. Huxor, {\it et~al.\/}, {\it Mon. Not. R. Astron. Soc.\/} {\bf 442}, 2165
  (2014).

\bibitem{Strader2011}
J.~Strader, N.~Caldwell, A.~C. Seth, {\it Astron. J.\/} {\bf 142}, 8 (2011).

\bibitem{Pritzl2005}
B.~J. Pritzl, K.~A. Venn, M.~Irwin, {\it Astron. J.\/} {\bf 130}, 2140 (2005).

\bibitem{Venn2004}
K.~A. Venn, {\it et~al.\/}, {\it Astron. J.\/} {\bf 128}, 1177 (2004).

\bibitem{Colucci2009}
J.~E. Colucci, R.~A. Bernstein, S.~Cameron, A.~McWilliam, J.~G. Cohen, {\it
  Astrophys. J.\/} {\bf 704}, 385 (2009).

\bibitem{Sakari2013}
C.~M. Sakari, M.~Shetrone, K.~Venn, A.~McWilliam, A.~Dotter, {\it Mon. Not. R.
  Astron. Soc.\/} {\bf 434}, 358 (2013).

\bibitem{Harris2006a}
W.~E. Harris, {\it et~al.\/}, {\it Astrophys. J.\/} {\bf 636}, 90 (2006).

\bibitem{Wan2020}
Z.~Wan, {\it et~al.\/}, {\it Nature\/} {\bf 583}, 768 (2020).

\bibitem{Usher2018}
C.~Usher, J.~Pfeffer, J.~M.~D. Kruijssen, R.~A. Crain, M.~Reina-Campos, {\it
  Mon. Not. R. Astron. Soc.\/} {\bf 480}, 3279 (2018).

\bibitem{ispy3}
S.~Larsen, {ISPy3: Integrated-light Spectroscopy for Python3,
  https://github.com/soerenslarsen/ISPy3, DOI 10.5281/zenodo.4036092} (2020).

\bibitem{Larsen1999}
S.~S. Larsen, {\it Astron. Astrophys. Suppl. Ser.\/} {\bf 139}, 393 (1999).

\bibitem{makee}
T.~Barlow, {MAKEE User Guide and Technical Documentation,
  https://www.astro.caltech.edu/{\~{}}tb/makee/} (2019).

\bibitem{Kroupa2001}
P.~Kroupa, {\it Mon. Not. R. Astron. Soc.\/} {\bf 322}, 231 (2001).

\bibitem{Kurucz1970}
R.~L. Kurucz, {Atlas: a Computer Program for Calculating Model Stellar
  Atmospheres,
  http://kurucz.harvard.edu/papers/sao309/saospecialreport309.pdf}, {\it Tech.
  rep.\/} (1970).

\bibitem{Sbordone2004}
L.~Sbordone, P.~Bonifacio, F.~Castelli, R.~L. Kurucz, {\it Memorie della
  Societ{\`{a}} Astronomica Italiana Supplement\/} {\bf 5}, 93 (2004).

\bibitem{kurucz}
R.~L. Kurucz, http://kurucz.harvard.edu (2020).

\bibitem{Kurucz1981}
R.~L. Kurucz, E.~H. Avrett, {Solar Spectrum Synthesis. I. A Sample Atlas from
  224 to 300 nm, http://kurucz.harvard.edu/papers/sao391/saosr391.pdf}, {\it
  Tech. rep.\/} (1981).

\bibitem{Dotter2007}
A.~Dotter, {\it et~al.\/}, {\it Astron. J.\/} {\bf 134}, 376 (2007).

\bibitem{Dotter2016}
A.~Dotter, {\it Astrophys. J. Suppl. Ser.\/} {\bf 222}, 8 (2016).

\bibitem{Salaris1993}
M.~Salaris, A.~Chieffi, O.~Straniero, {\it Astrophys. J.\/} {\bf 414}, 580
  (1993).

\bibitem{Pietrinferni2006}
A.~Pietrinferni, S.~Cassisi, M.~Salaris, F.~Castelli, {\it Astrophys. J.\/}
  {\bf 642}, 797 (2006).

\bibitem{Cox2000}
A.~N. Cox, {\it {Allen's Astrophysical Quantities}\/} (AIP Press, New York,
  2000), fourth edn.

\bibitem{Kurucz2005}
R.~L. Kurucz, {\it Memorie della Societ{\`{a}} Astronomica Italiana
  Supplement\/} {\bf 8}, 14 (2005).

\bibitem{McWilliam2008}
A.~McWilliam, R.~A. Bernstein, {\it Astrophys. J.\/} {\bf 684}, 326 (2008).

\bibitem{Eitner2019}
P.~Eitner, M.~Bergemann, S.~S. Larsen, {\it Astron. Astrophys.\/} {\bf 627},
  A40 (2019).

\bibitem{Castelli2005}
F.~Castelli, {\it Memorie della Societ{\`{a}} Astronomica Italiana
  Supplement\/} {\bf 8}, 44 (2005).

\bibitem{hireswww}
{HIRES Home Page, https://www2.keck.hawaii.edu/inst/hires/}.

\bibitem{King1962}
I.~King, {\it Astron. J.\/} {\bf 67}, 471 (1962).

\bibitem{Stanek1998}
K.~Z. Stanek, P.~M. Garnavich, {\it Astrophys. J.\/} {\bf 503}, L131 (1998).

\bibitem{Larsen2002b}
S.~S. Larsen, J.~P. Brodie, A.~Sarajedini, J.~P. Huchra, {\it Astron. J.\/}
  {\bf 124}, 2615 (2002).

\end{thebibliography}

\bibliographystyle{Science}

\begin{figure}
\includegraphics[width=15cm]{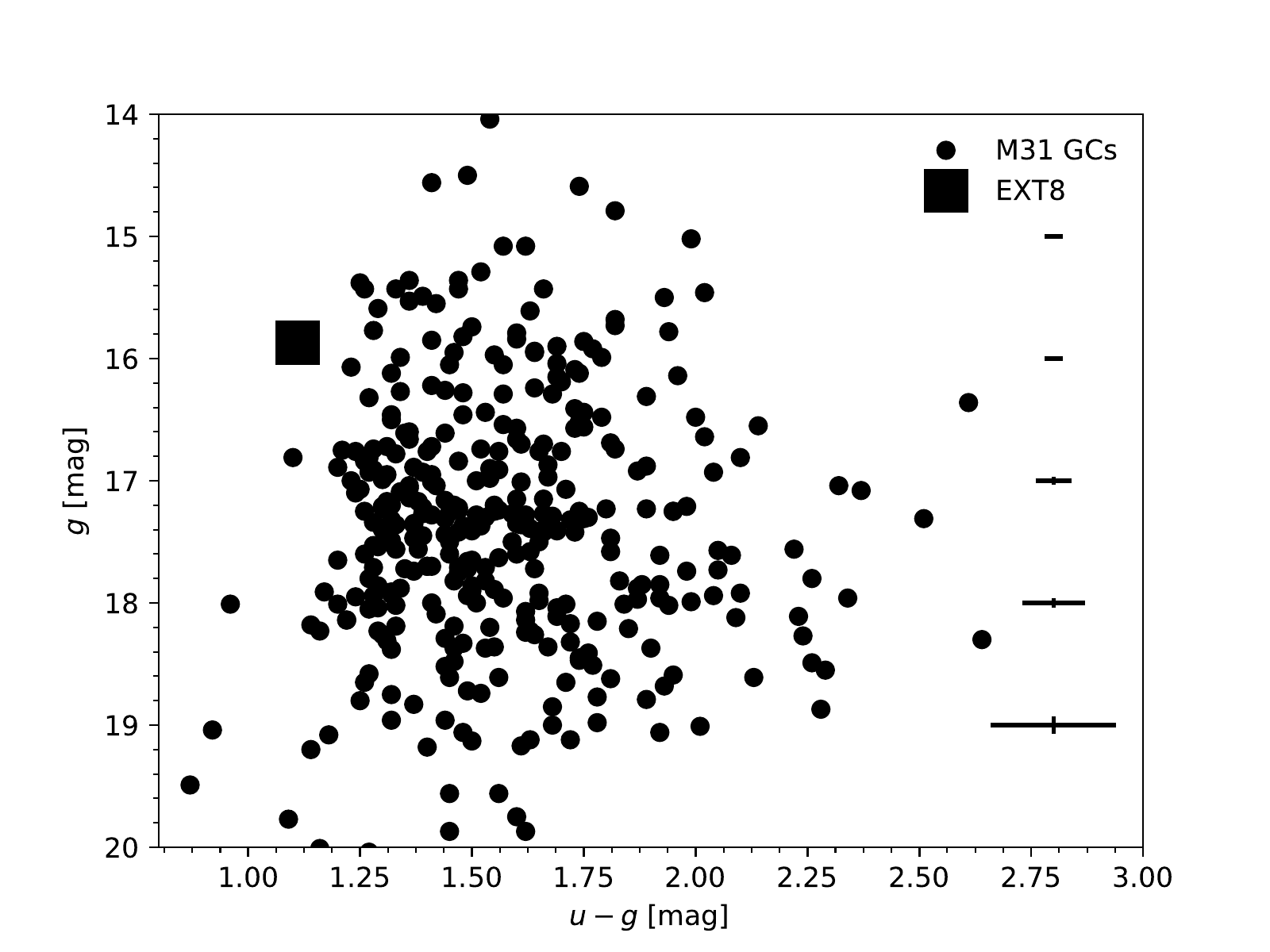}
\caption{\textbf{Color-magnitude diagram for globular clusters in M31 \cite{Peacock2010}}. No correction for dust reddening has been applied. EXT8 is marked with a large square, and has one of the bluest $u-g$ colors among the GCs in M31. Typical one-sigma error bars are shown on the right.}
\end{figure}

\begin{figure}
\includegraphics[width=10cm]{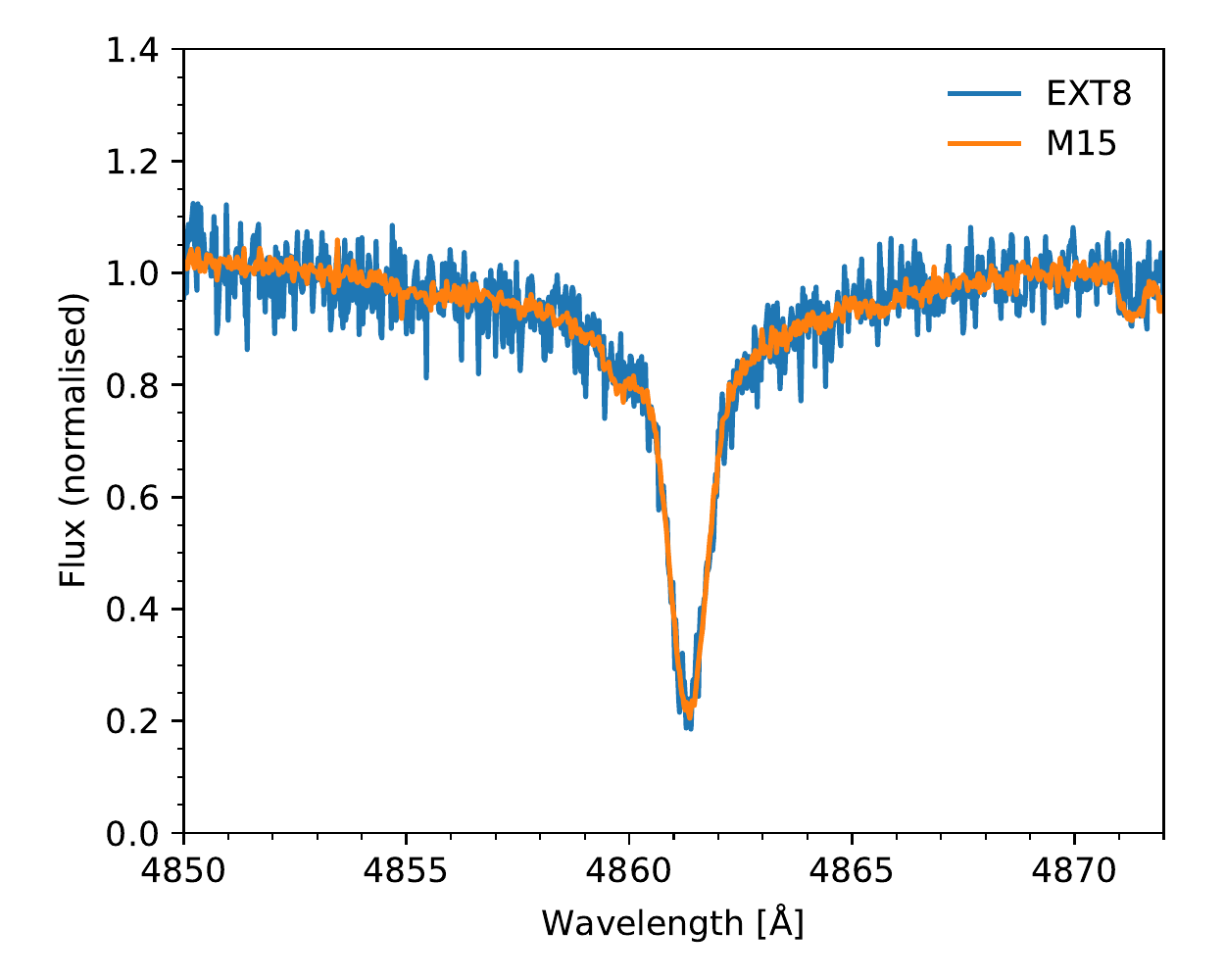}
\caption{\textbf{Comparison of the H$\beta$ lines in the spectra of EXT8 (blue) and Messier 15 (orange).} The very similar H$\beta$ line profiles in the two spectra indicate similar, old ages for the two clusters. }
\end{figure}

\begin{figure}
\includegraphics[width=16cm]{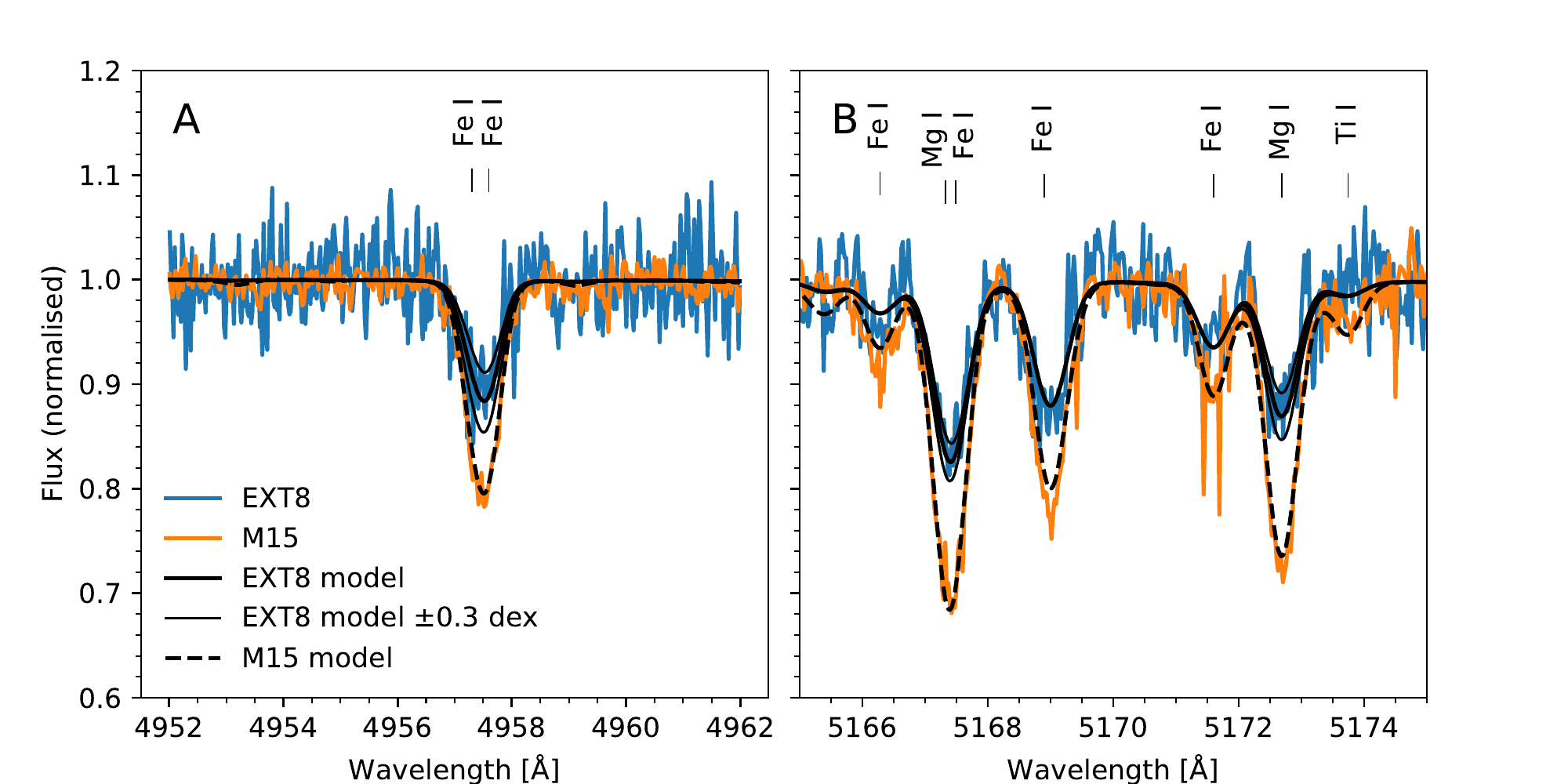}
\caption{\textbf{Iron- and magnesium features in the spectra of EXT8 (blue) and Messier~15 (orange)}. The  best-fitting models are shown with thick black lines (for EXT8) and dashed black lines (for M15). Models for EXT8 in which the abundances have been varied by $\pm0.3$ dex for iron (panel A) and magnesium (panel B) are shown with thinner lines.}
\end{figure}

\begin{figure}
\includegraphics[width=16cm]{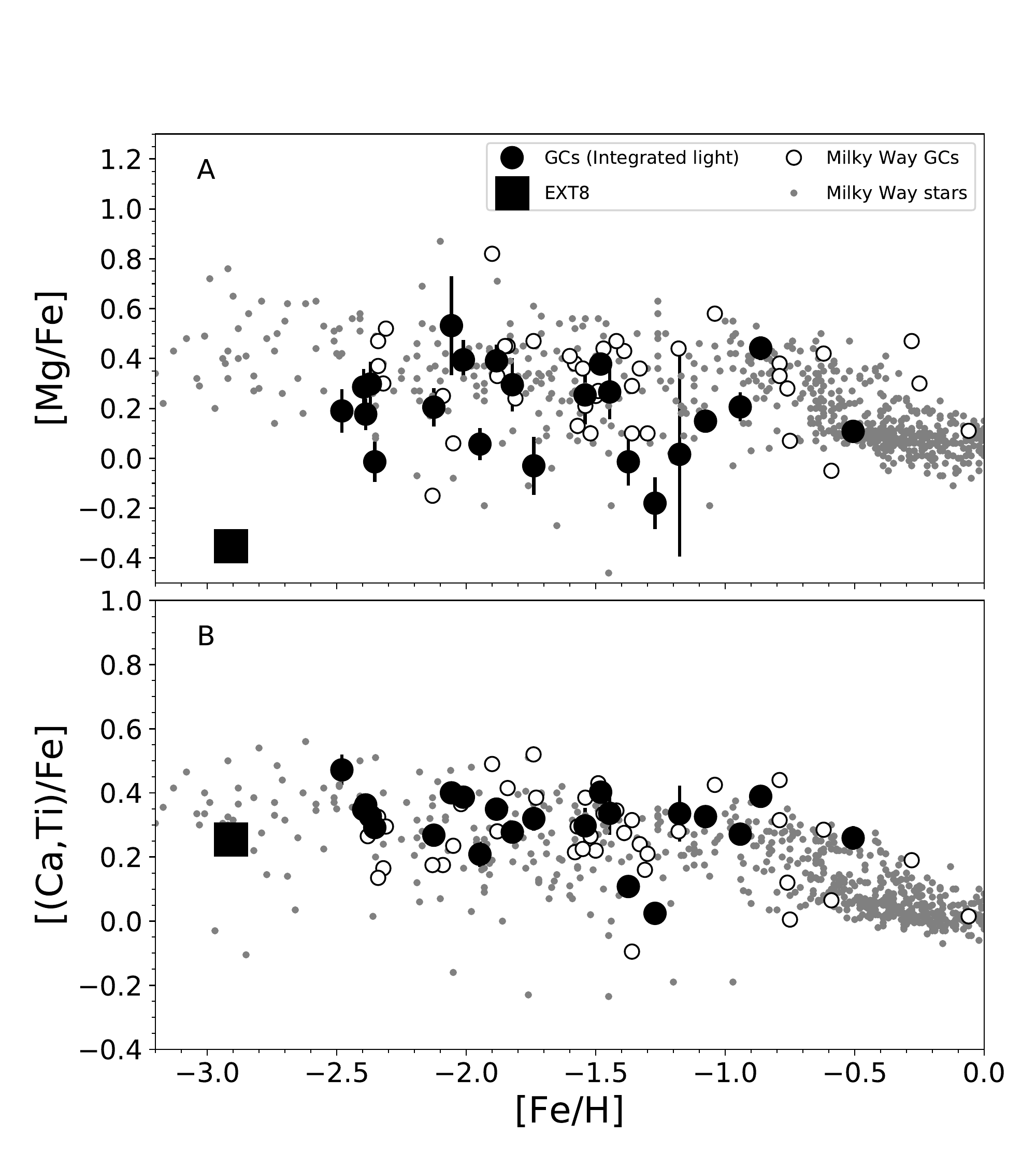}
\caption{\textbf{Abundance measurements for EXT8 and other Galactic and extragalactic GCs}. 
Panel (A) shows $\mathrm{[Mg/Fe]}$ as a function of $\mathrm{[Fe/H]}$ and Panel (B) shows the average of $\mathrm{[Ca/Fe]}$ and $\mathrm{[Ti/Fe]}$. 
The large square shows our  measurements for EXT8 while the filled circles show integrated-light measurements for GCs in the NGC~147, NGC~6822, Fornax, and WLM dwarf galaxies, M~33, and the Milky Way \cite{Larsen2017,Larsen2018}.  Data for resolved Milky Way GCs are shown with open circles \cite{Pritzl2005} and data for field stars \cite{Venn2004} with small gray dots. Error bars specify the one sigma uncertainties.}
\end{figure}

\section*{Acknowledgments}
We thank Mark Gieles and Else Starkenburg for helpful discussions and comments on the manuscript, and
Don VandenBerg for valuable advice on the selection of isochrones. 
Comments from the anonymous referees helped improve the presentation. 
The data presented herein were obtained at the W. M. Keck Observatory, which is operated as a scientific partnership among the California Institute of Technology, the University of California and the National Aeronautics and Space Administration. The Observatory was made possible by the generous financial support of the W. M. Keck Foundation. The authors wish to recognize and acknowledge the very significant cultural role and reverence that the summit of Maunakea has always had within the indigenous Hawaiian community.  We are most fortunate to have the opportunity to conduct observations from this mountain.

\paragraph{Funding:} AJR was supported by National Science Foundation grant AST-1616710, and as a Research Corporation for Science Advancement Cottrell Scholar. JPB acknowledges support from HST grant HST-GO-15078.
AJR and SSL were supported in part by the National Science Foundation under Grant No. NSF PHY-1748958

\paragraph{Authors contributions:}JPB secured the observing time for this project, all authors contributed to the planning of the observations, and the inclusion of EXT8 as a target was suggested by AJR. AW conducted the observations and SSL carried out the data reduction and analysis and drafted the paper. All authors assisted in the interpretation of the results and writing of the paper. 

\paragraph{Competing interests:} None.

\paragraph{Data and materials availability:}
The average measured abundances are listed in Table~S2 and individual measurements are in Tables S3-S9. The raw spectra of EXT8 and Hodge~III are available in the W. M. Keck Observatory Archive at http://koa.ipac.caltech.edu, program ID U177Hr, Semester 2019B, and program ID U040Hr, Semester 2015B, P.I. Brodie.
The UVES observations of M15 are available in the ESO archive at http://archive.eso.org, program ID 095.B-0677(A), P. I. Larsen. 
The CFHT MegaCam images of EXT8 are available in the CFHT data archive at http://www.cadc-ccda.hia-iha.nrc-cnrc.gc.ca/en/cfht, proposal ID 16BC25, product IDs 2011389p, 2011390p, and 2011391p. 
The spectral analysis was carried out with the \textsc{ISPy3} code \cite{ispy3}, and the King model fitting was done with the \textsc{Baolab} code \cite{Larsen1999}, available at https://github.com/soerenslarsen/baolab, DOI 10.5281/zenodo.4036106. 

\nocite{makee}
\nocite{Kroupa2001}
\nocite{Kurucz1970}
\nocite{Sbordone2004}
\nocite{kurucz}
\nocite{Kurucz1981}
\nocite{Dotter2007}
\nocite{Dotter2016}
\nocite{Salaris1993}
\nocite{Pietrinferni2006}
\nocite{Cox2000}
\nocite{Kurucz2005}
\nocite{McWilliam2008}
\nocite{Eitner2019}
\nocite{Castelli2005}
\nocite{hireswww}
\nocite{King1962}
\nocite{Stanek1998}
\nocite{Larsen2002b}

\section*{Supplementary materials}
Materials and Methods\\
Supplementary Text\\
Figures S1-S3\\
Tables S1 to S10 \\
References (\textit{39-57})

\end{document}